\documentclass[preprint2]{aastex}

\begin{document}

\title{GAS EMISSION SPECTRUM IN THE~Irr~GALAXY IC~10}

\author{T.A. Lozinskaya \altaffilmark{1},
O.V. Egorov \altaffilmark{1},
A.V. Moiseev \altaffilmark{2},
D.V. Bizyaev \altaffilmark{1,3}
}

\altaffiltext{1}{Sternberg Astronomical Institute,
Universitetskiy pr. 13, Moscow, 119992 Russia}

\altaffiltext{2}{Special Astrophysical Observatory, Russian Academy of
Sciences, Nizhniy Arkhyz, Karachai-Cherkessian Republic, 357147 Russia}

\altaffiltext{3}{New Mexico State University and
Apache Point Observatory, Sunspot, NM, USA}

\begin{abstract}
Spectroscopic long-slit observations of the dwarf Irr~galaxy IC~10
were conducted at the 6-m Special Astrophysical Observatory
telescope with the SCORPIO focal reducer. The ionized-gas
emission spectra in the regions of intense current star formation
were obtained for a large number of regions in IC~10. The relative
abundances of oxygen, N$^+$, and S$^+$ in about twenty
HII~regions and in the synchrotron superbubble were estimated. We
found that the galaxy-averaged oxygen abundance is
$12+\log(\textrm{O/H}) = 8.17 \pm 0.35$ and the metallicity is
$Z=0.18 \pm 0.14 Z_{\odot} $. Our abundances estimated from the
strong emission lines are found to be more reliable than those
obtained by comparing diagnostic diagrams with photoionization
models.

\end{abstract}

\keywords{Irr~galaxies, IC~10, interstellar medium, chemical
composition.}

\maketitle

\section*{INTRODUCTION}

The dwarf Irr~galaxy IC~10 is a unique testing ground for
investigating the interstellar medium in regions of violent star
formation. The galaxy's images in the~H$\alpha$ and~[SII] lines
appear as a single giant complex of multiple shells and
supershells with sizes from~50 to~800--1000~pc (Zucker~2000;
Wilcots and Miller~1998; Gil~de~Paz et~al.~2003; Leroy
et~al.~2006; Chyzy et~al.~2003; Lozinskaya et~al.~2008). Its
stellar population suggests a recent starburst (t=4 -- 10~Myr)
and an older starburst ($t>350$~Myr) (Hunter~2001; Zucker~2002;
Massey et~al.~2007; Vacca et~al.~2007; and references therein).
The high H$\alpha$ and infrared luminosities and the anomalously
large number of Wolf--Rayet (WR) stars in IC~10 (the WR space
density is highest among the dwarf galaxies, comparable to that
in massive spiral galaxies (Massey et~al.~1992,~2007; Richer
et~al.~2001; Massey and Holmes~2002; Crowther et~al.~2003; Vacca
et~al.~2007)) are indicative of a short current starburst
affecting the bulk of the galaxy.

\begin{table*}[t!]
\begin{center}
\caption{Summary of observations}\medskip
\begin{tabular}{llcccc}
\hline \multicolumn{1}{c}{Slit (PA)} & Date &$\Delta\lambda$, \AA&
$\delta\lambda$, \AA  & $T_{\textrm{exp}}$, s & Seeing, arcsec \\
\hline
PA0      & Aug.~17/18, 2006& 3620--5370 & 5\phantom{0.}  &6000 & 2.5 \\
PA0      & Sep.~2/3, 2008& 5650--7340 & 5\phantom{0.}  &3600 & 2.1 \\
PA132    & Aug.~17/18, 2006& 3620--5370 & 5\phantom{0.}  &4800 & 1.4 \\
PA132    & Aug.~18/19, 2006& 5650--7340 & 5\phantom{0.}  &3600 & 1.4 \\
PA268    & Feb.~14/15, 2007& 6060--7060 & 2.5            &1200 & 1.4 \\
PA268    & Jan.~15/16, 2008& 3620--5370 & 5\phantom{0.}  &4800 & 1.5 \\
PA45     & Oct.~29/30, 2008& 3650--7350 & 9\phantom{0.}  &3600 & 1.0 \\
PA331    & Oct.~28/29, 2008& 3650--7350 & 9\phantom{0.}  &6000 & 0.9 \\
\hline
\end{tabular}
\end{center}
\end{table*}

Lozinskaya and Moiseev~(2007) explained the formation of the unique
synchrotron
superbubble (SS), which was previously associated with multiple supernova
explosions
(Yang and Skillman~1993; Bullejos and Rozado~2002; Rozado et~al.~2002;
Thurow
and Wilcots~2005), by a hypernova explosion.

In this paper we investigate emission spectra of the ionized gas in the
regions of current star formation in IC~10 based on observations
conducted with the 6-m
Special Astrophysical Observatory (SAO) telescope and SCORPIO focal
reducer operating in the mode of a long-slit spectrograph. The main goal of our
observations is to estimate the metallicity for a large number of emission
regions. Previously, Lequeux et~al.~(1979), Garnett~(1990), Richer
et~al.~(2001), and Lee et~al.~(2003) determined the metallicity only in
three brightest HII~regions of the galaxy.

In the succeeding sections, we describe the observing and data reduction
techniques, present and discuss the results obtained, and, in conclusion,
summarize our main conclusions.

\section*{OBSERVATIONS AND DATA REDUCTION}

The observations of IC~10 were performed with the SCORPIO
instrument (Afanasiev and Moiseev~2005) operating in the mode of
a long-slit spectrograph with a slit about~$6'$ in length and~$1''$ in
width. The scale along the slit was 0.36$''$ per~pixel.

For five slit positions, we obtained spectra with a resolution from~2.5 to
10~\AA. Below, the spectrograms are designated as PA0, PA45, PA132, PA268, and
PA331, in accordance with their position angles. The corresponding spectral
range~$\Delta\lambda$, spectral resolution~$\delta\lambda$, total exposure
time~$T_{\textrm{exp}}$, and average seeing are given in Table~1.

The data were reduced in a standard way; the spectra of the stars BD+25d4655,
GRW+70d5824, and GD~71 observed immediately after the object at the same zenith
distance were used to calibrate the energy scale.

In the spectra being analyzed, we determined the continuum level from the
underlying stellar population based on spline fitting. The integrated
emission-line flux was measured by means of single-component Gaussian fitting.
To increase the signal-to-noise ratio for faint emission regions, we performed
an averaging over individual HII~regions when processing the spectrograms, with
the domain of integration being from~$2''$ to~$20''$ in size. The errors in the
line fluxes were estimated from the synthetic spectra that simulated
observations with the required signal-to-noise ratio. The ranges of errors
given below in tables and figures correspond to~$3\sigma$.

\section*{RESULTS OF OBSERVATIONS}

\subsection*{The Ionized-Gas Emission Spectrum}

The positions of five slits passing through the synchrotron superbubble, the
bright region of current star formation, and fainter shell structures are shown
in Figs.~1a and~1b. The spectroscopically confirmed WR stars~from the lists by
Royer et~al.~(2001) and Massey and Holmes~(2002) are also marked there. For the
convenience of identification, the slit spectrograms in the figures are marked
up in arcsec.

The measurements performed at the points of intersection between spectrograph
slits allowed the actual accuracy of our observations to be estimated. Table~2
lists the relative line fluxes measured from a pair of slit spectrograms
at their intersections. As we see, within the error limits, the agreement is
largely good. The differences exceeding the formal observational errors may be
related to different domains of integration of the fluxes for two slit
positions on the nebula with a fine structure.

\begin{figure*}[p!]
\begin{center}
\includegraphics[scale=0.75]{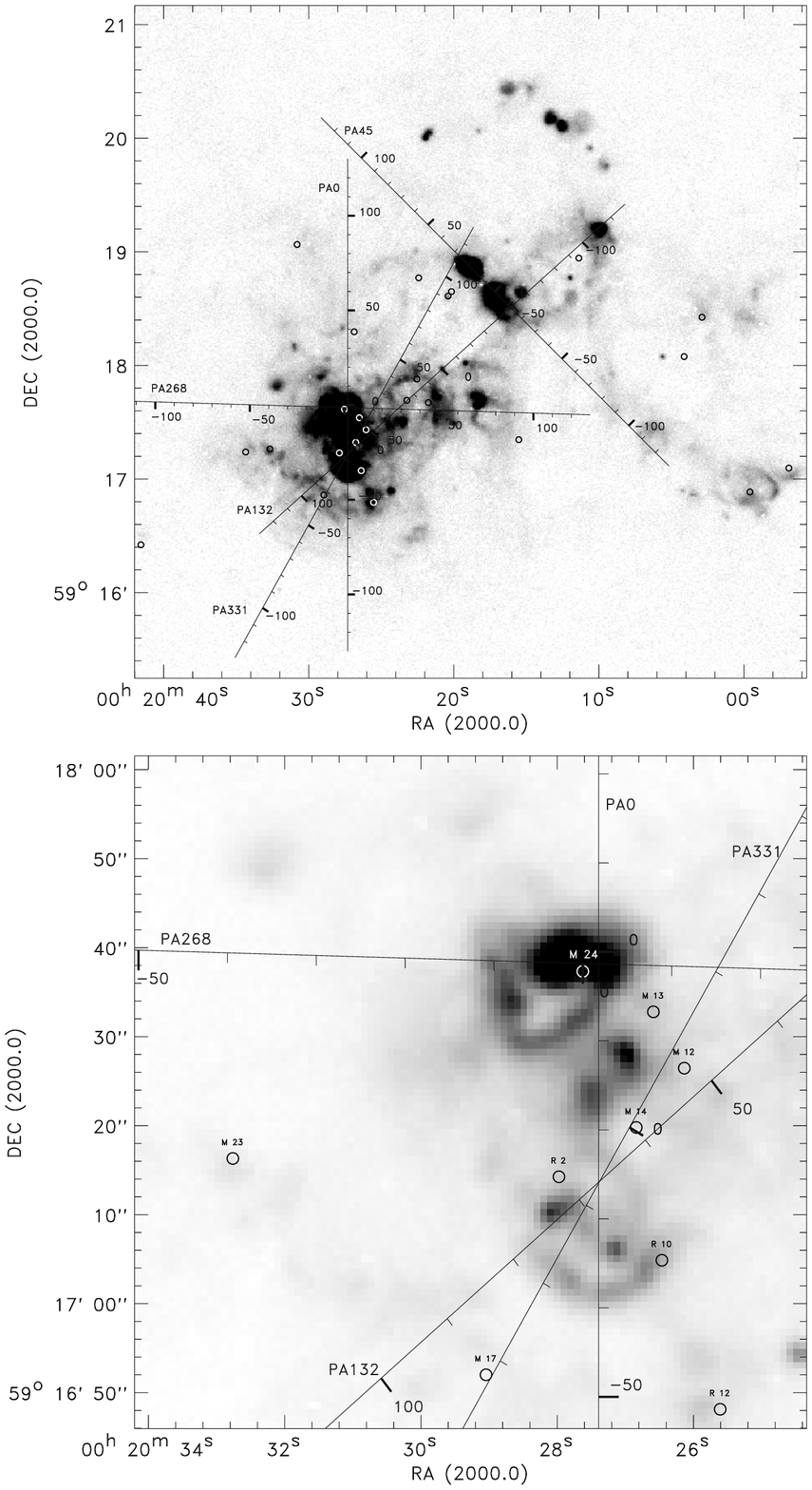}
\end{center}

\vspace*{10pt} \caption{Positions of our five long slits
on the H$\alpha$ image of IC~10:
(a)~the entire galaxy and~(b) the bright region of current
star formation. The long-slit spectrograms are marked up in arcsec.
The circles designate the spectroscopically confirmed WR stars from
Royer et~al.~(2001)
and Massey and Holmes~(2002). \hfill}
\end{figure*}

\begin{figure*}[t!]

\includegraphics[scale=0.75]{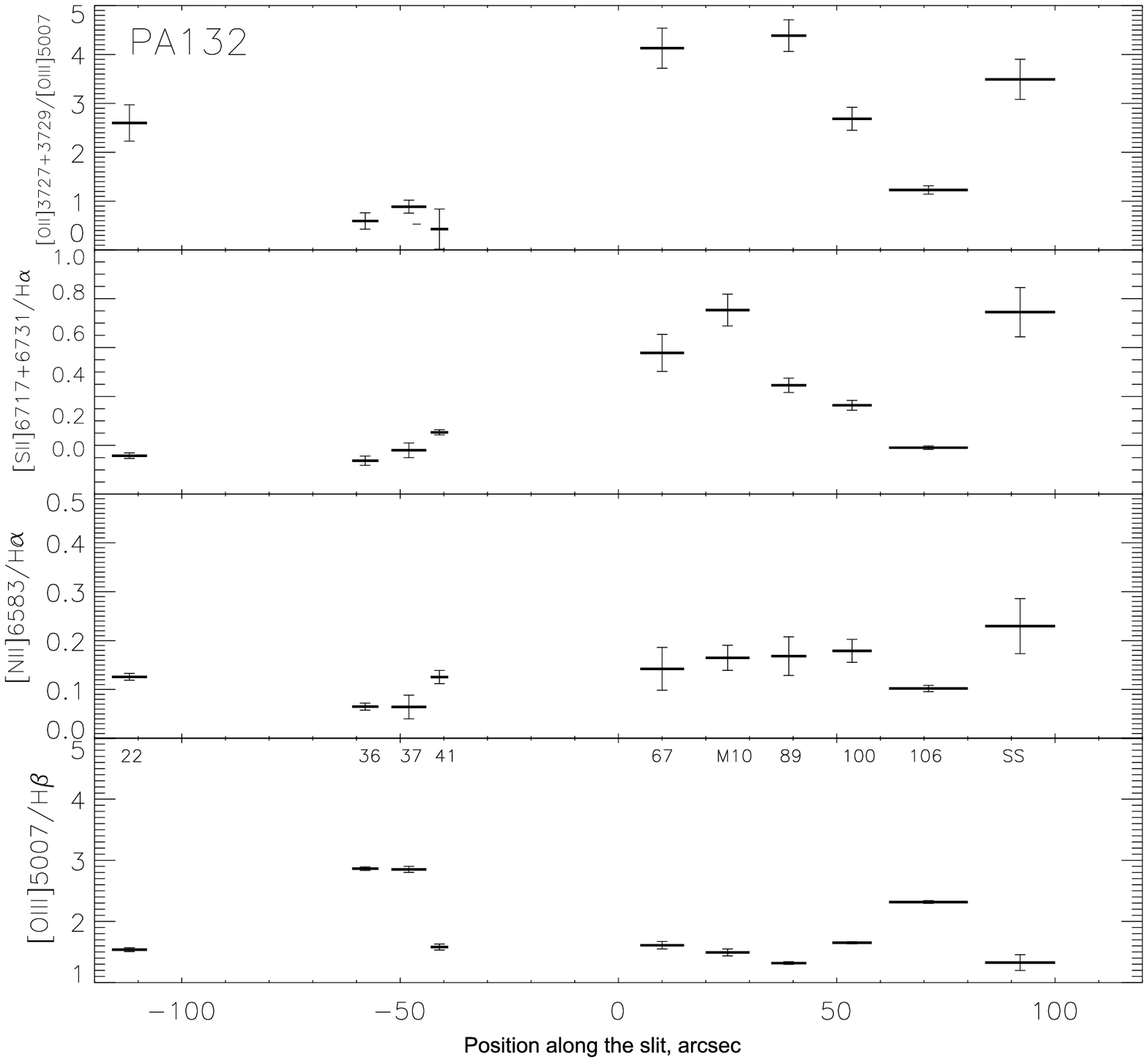}
\caption{Distributions of the line flux ratios
$I([\textrm{OII}]\lambda3727+3729\textrm{\AA})/I([\textrm{OIII}]\lambda5007\textrm{\AA})$,
$I([\textrm{SII}]\lambda6717+6731\textrm{\AA})/I(\textrm{H}\alpha)$,
$I([\textrm{NII}]\lambda6583\textrm{\AA})/I(\textrm{H}\alpha)$,
and $I([\textrm{OIII}]\lambda5007\textrm{\AA})/I(\textrm{H}\beta)$
along slit~PA132.
Positions along the slits are given in arcsec at the
horizontal axis in the lower panel. The numbers at the top of the
lower panel denote the HII~regions from the catalog by Hodge and
Lee~(1990). "M10" marks the faint region around the WR star~M10, and
"SS" designates the synchrotron superbubble. \hfill}

\end{figure*}

\begin{figure*}[t!]
\includegraphics[scale=0.80]{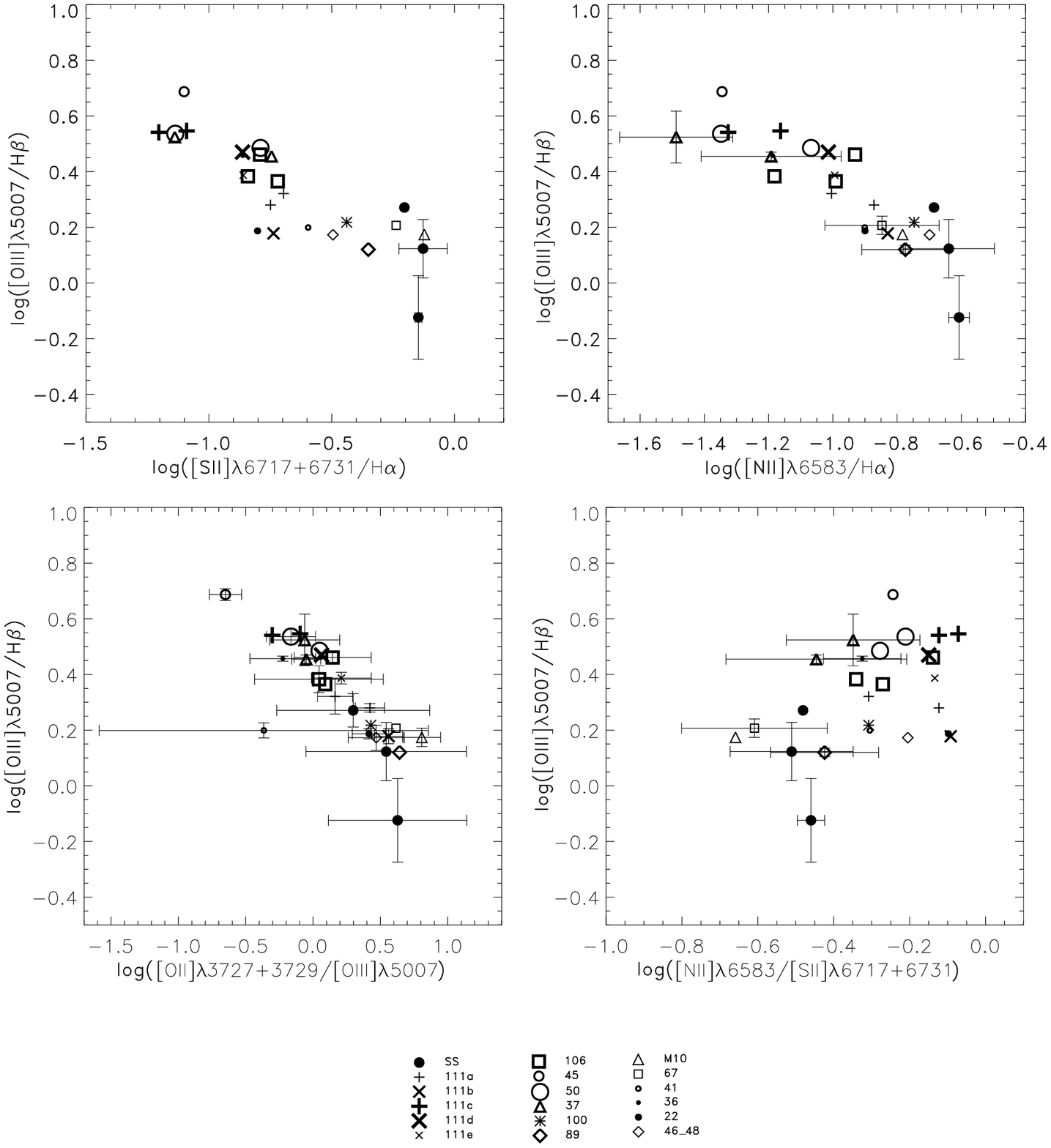}
\caption{Diagnostic diagrams of the line flux ratios:
$I([\textrm{OIII}]\lambda 5007\textrm{\AA})/I(\textrm{H}\beta)$
versus
$I([\textrm{SII}]\lambda6717+6731\textrm{\AA})/I(\textrm{H}\alpha)$,
versus
$I([\textrm{NII}]\lambda6583\textrm{\AA})/I(\textrm{H}\alpha)$,
versus
$I([\textrm{OII}]\lambda3727+3729\textrm{\AA})/I([\textrm{OIII}]\lambda5007\textrm{\AA})$,
and versus
$I([\textrm{NII}]\lambda6583\textrm{\AA})/I([\textrm{SII}]\lambda6717+6731\textrm{\AA}).$
The different symbols indicated at the bottom give the relative
intensities averaged over individual HII~regions from the catalog
by Hodge and Lee~(1990), over the region around~WR~M10, and over
the synchrotron superbubble~(SS).
The uncertainties that exceed 0.1 dex are shown as the error bars. \hfill}
\end{figure*}

\begin{figure*}[t!]
\includegraphics[scale=0.85]{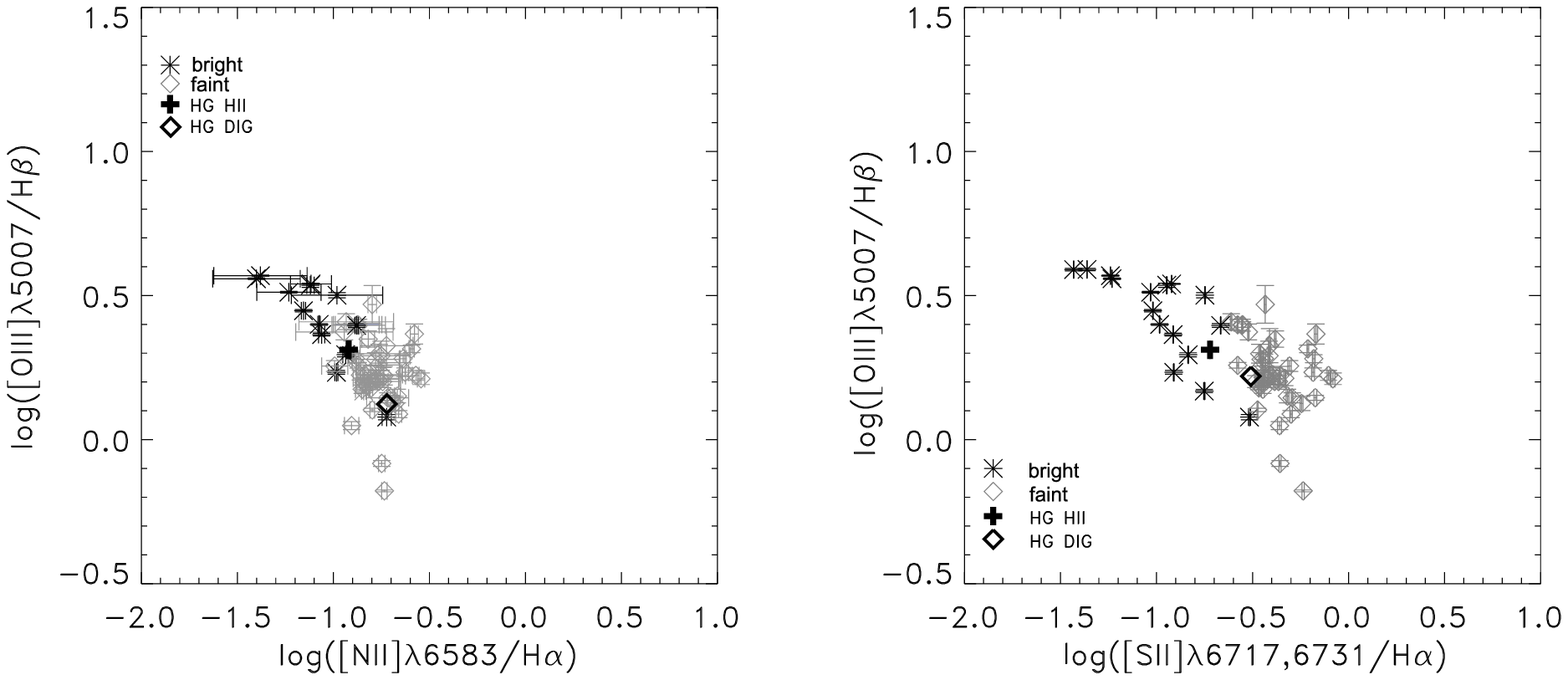}

\caption{Diagnostic diagrams with bright
($F_{\textrm{H}\alpha} \geq 1.0 \cdot
10^{-15}$~erg~s$^{-1}$~cm$^{-2}$, black asterisks) and faint
($F_{\textrm{H}\alpha} \leq 1.0\cdot
10^{-15}$~erg~s$^{-1}$~cm$^{-2}$, gray diamonds) regions. The
measurements in 5-pixel ($1''.8$) fields along the slits~PA0, PA132,
and~PA268 are presented. The heavy cross and diamond mark the
galaxy-averaged values found by Hidalgo-Gamez~(2005) for bright
HII~regions and the diffuse ionized gas (DIG). \hfill}
\end{figure*}

\begin{figure*}[t!]
\begin{center}
\includegraphics[scale=0.70]{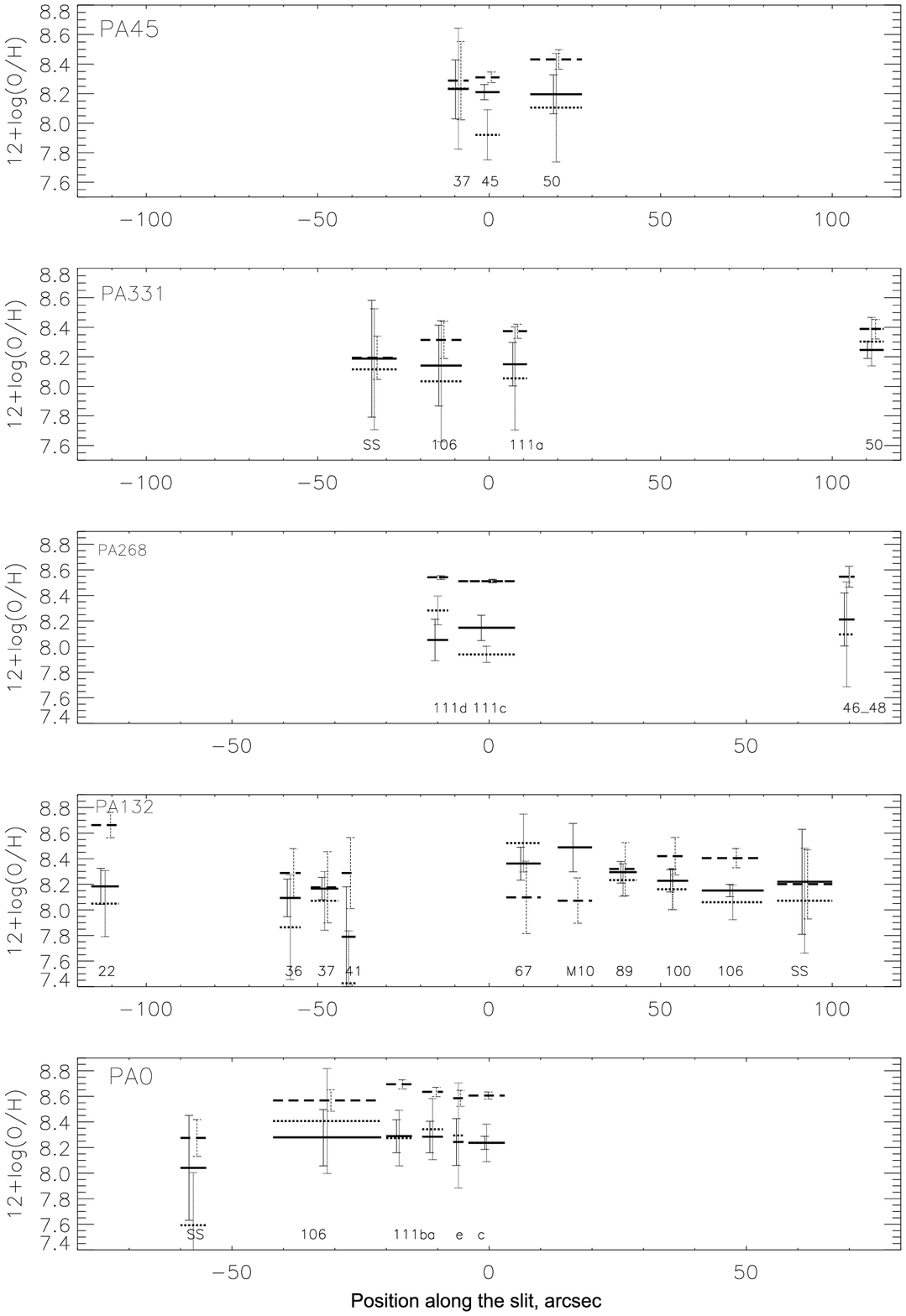}
\end{center}
\caption{Distributions of abundances $12+ \log(\textrm{O/H})$~(a),
$\log(\textrm{N}^{+}/\textrm{O})$~(b), and
$\log(\textrm{S}^{+}/\textrm{O})$~(c) along our five slits.
The HII~region identifications are
signed at the horizontal axis of the figures. The oxygen
abundances $12+ \log(\textrm{O/H})$ found using the methods by
Izotov et~al.~(2006), Pilyugin and Thuan~(2005), and Charlot and
Longhetti~(2001) are shown by dots with the solid, dotted, and dashed
error bars, respectively. \hfill}

\end{figure*}

\begin{figure*}[t!]
\begin{center}
\includegraphics[scale=0.70]{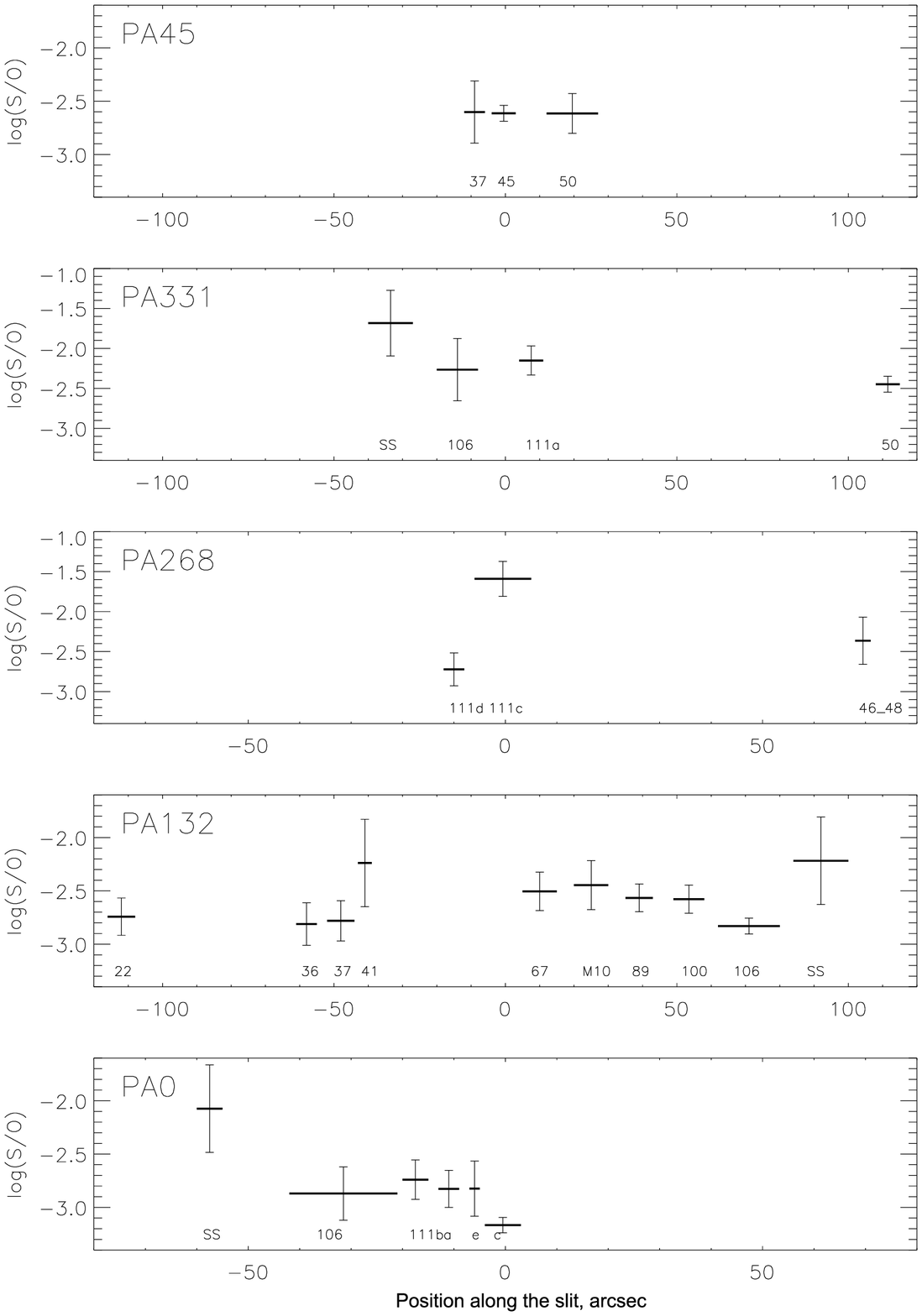}
\end{center}

\addtocounter{figure}{-1}
 \vspace*{5pt}\caption{(Contd.) \hfill}
\end{figure*}

\begin{figure*}[t!]
\begin{center}
\includegraphics[scale=0.70]{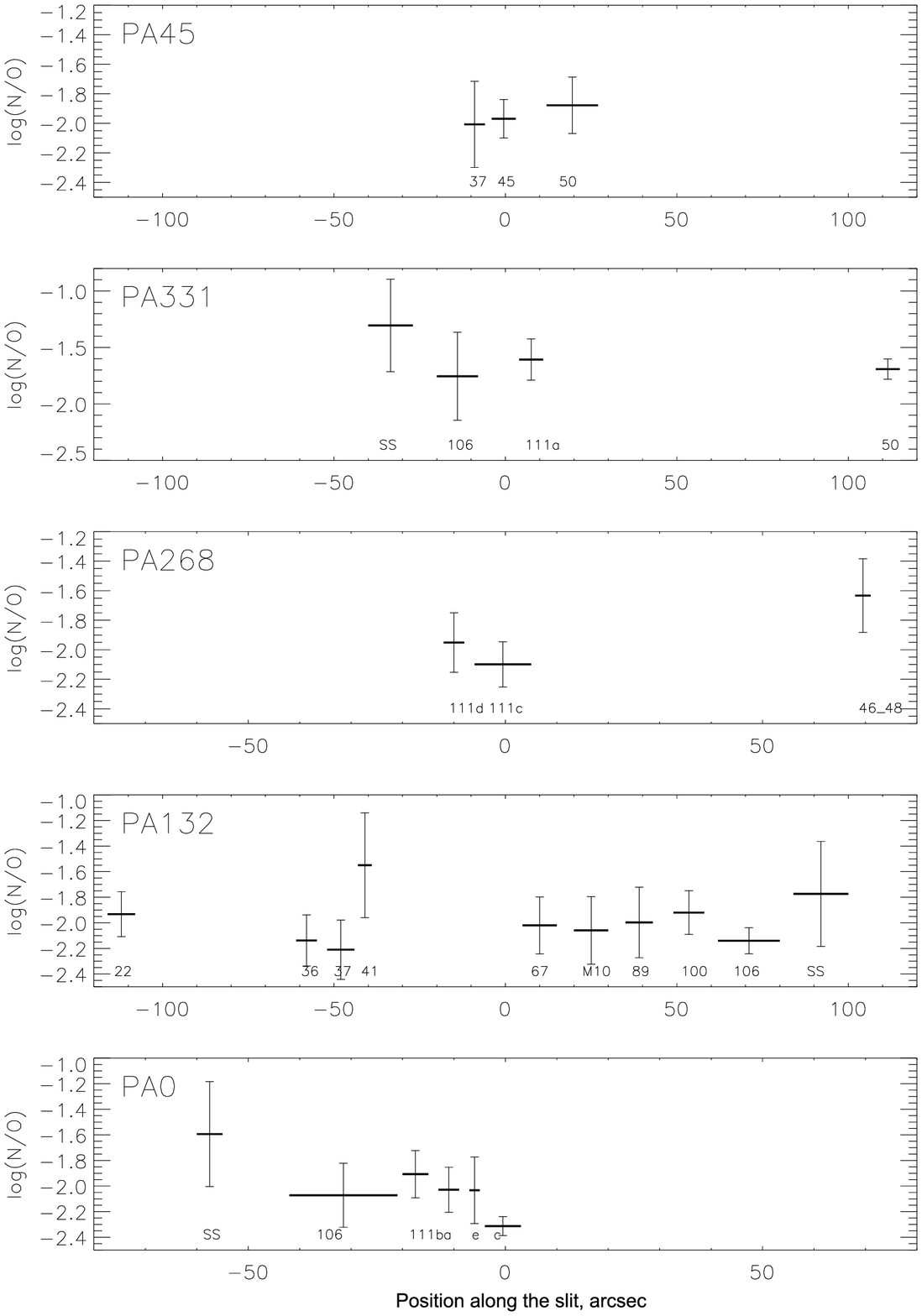}
\end{center}
\vspace*{5pt} \addtocounter{figure}{-1}

\caption{(Contd.) \hfill}
\end{figure*}

Results of the spectra reduction are presented in Fig.~2 and
Table~3. Fig. 2 shows the distributions of the relative line
intensities $I$([OII]$\lambda3727+3729$\AA) /
$I$([OIII]$\lambda5007$\AA), $I$([SII]$\lambda6717+6731$\AA) /
$I$(H$\alpha$), $I$([NII]$\lambda6583$\AA) / $I$(H$\alpha$), and
$I$([OIII]$\lambda$5007\AA) / $I$(H$\beta$) along slit~PA132. The
slit is marked up in arcsec along the horizontal axis in the lower
panel, the numbers at the top denote the HII~regions from the
catalog by Hodge and Lee~(1990). "M10" designates the faint region
around the WR star~M10 unrecorded in the catalog; SS designates
the synchrotron superbubble. As it is seen in Fig.~1, the
slit~PA331 (positions~4--11) passes over the faint region near
the~HL111a in the southeast and positions~108--115 correspond to
the northern periphery of the region~HL50. To avoid overloading
the figures and tables, these two regions on slit~PA331 are
designated in Tables~3 and~4 and all figures simply as~HL111a
and~HL50, respectively. The slit~PA268 (positions~68--71) passes
between the bright nebulae~HL46 and~HL48; this region is
designated below as~HL46\_48.

The columns in Table~3 show the designation of HII~regions from
the catalog by Hodge and Lee~(1990), our long-slit spectrogram
utilized for the estimations and the used positions on it, the
relative intensities of the above-mentioned emission lines, and
the [SII]$\lambda6717/6731$~\AA\ doublet line intensity ratios.

The interstellar extinction in Table~3 and Fig.~2 was taken into
account. We determined the color excess
$E(B - V)$ individually for each HII~region, where possible. The most
reliable results were obtained from the low-spectral-resolution
spectrograms~PA45 and~PA331. For these spectrograms, we made our estimates by
comparing the observed H$\alpha$:H$\beta$:H$\gamma$ flux ratio with the
theoretical one: H$\alpha$:H$\beta$:$H\gamma  = 2.86:1.00:0.47$, which,
according to Aller~(1984), is valid for a typical electron density in a
star-forming region of 30--300~cm$^{-3}$ when
the electron temperature is accepted to be 11000~K
(see discussion below). For the regions that did not fall in these slits,
we assumed either
the mean value of $E(B-V)$ = 0.95 mag  or the color excess for
the close region that fell into the slits PA~45 or PA~331.
Our estimates of $E(B - V)$ for different regions in IC~10
range between 0.8 and 1.1 mag.

\subsection*{Diagnostic Diagrams of Relative Line Intensities}

Fig.~3 shows the diagnostic diagrams of relative line intensities for IC~10
that are traditionally used to compare observations with computed ionization
models: $I([\textrm{OIII}]\lambda 5007$\AA) / $I(\textrm{H}\beta)$ versus
$I$([SII]$\lambda6717+6731$\AA) / $I(\textrm{H}\alpha)$, versus
$I$([NII]$\lambda6583$\AA) / $I(\textrm{H}\alpha)$, versus
$I$([OII]$\lambda3727+3729$\AA) / $I([\textrm{OIII}]\lambda5007\textrm{\AA})$,
and versus
$I$([NII]$\lambda6583$\AA) / $I([\textrm{SII}]\lambda6717+6731\textrm{\AA})$.
Different symbols indicate the relative intensities averaged over individual
HII~regions from the catalog by Hodge and Lee~(1990), over the region around
WR~M10, and over the synchrotron superbubble~(SS).
The observing uncertainties greater than 0.1 dex are shown as the error bars
in Fig.~3.

In Figs.~4a and~4b, the data were combined separately by bright
($F_{\textrm{H}\alpha} \geq 1.0\times
10^{-15}$~erg~s$^{-1}$~cm$^{-2}$) and faint
($F_{\textrm{H}\alpha} \leq 1.0\times
10^{-15}$~erg~s$^{-1}$~cm$^{-2}$) emission regions. The measurements along
the 1.8" wide slits~PA0, PA132, and PA268 are shown in Fig.~4.
The crosses and diamonds in these figures also indicate the
galaxy-averaged values found by Hidalgo-Gamez~(2005) for bright
HII~regions and faint regions of diffuse ionized gas (DIG) (see
the ``Discussion'' section).

\subsection*{Estimates of the Relative Oxygen, Nitrogen, and Sulfur Abundances}

The spectra enable us to estimate the
relative oxygen abundance in a large number of galactic
HII~regions, both bright and faint. Previously, such estimates
were made for only three brightest regions: HL111, HL45, and the compact clump
HL106a inside HL106 (Lequeux et~al. 1979; Garnett~1990; Richer et~al. 2001; Lee
et~al. 2003).

\begin{table*}[t!]
\begin{center}

\caption{Emission line flux ratios at the points of intersection of our slit
spectrograms}\medskip
\begin{tabular}{lcc}
\hline \multicolumn{1}{c}{Spectrogram} &
$\log$($I([\textrm{OIII}]\lambda5007~\textrm{\AA})/I(\textrm{H}\beta)$) &
$\log$($I([\textrm{SII}]\lambda6717+6731~\textrm{\AA})/I(\textrm{H}\alpha)$)\\
\hline
& \multicolumn{2}{l}{1. Intersection of PA0, PA132, PA331}\\
PA0   & $0.32 \pm 0.03$ & $-0.86 \pm 0.02$ \\
PA132 & $0.30 \pm 0.03$ & $-0.68 \pm 0.04$ \\
PA331 & $0.31 \pm 0.03$ & $-0.84 \pm 0.03$\\
& \multicolumn{2}{l}{2. Intersection of PA0, PA268}\\
PA0   & $0.55 \pm 0.03$ & $-1.10 \pm 0.04$ \\
PA268 & $0.55 \pm 0.02$ & $-1.25 \pm 0.03$ \\
& \multicolumn{2}{l}{3. Intersection of PA268, PA331}\\
PA268 & $0.22 \pm 0.04$ & $-0.45 \pm 0.03$ \\
PA331 & $0.18 \pm 0.05$ & $-0.47 \pm 0.04$ \\
& \multicolumn{2}{l}{4. Intersection PA132, PA268}\\
PA132 & $0.02 \pm 0.04$ & $-0.18 \pm 0.08$ \\
PA268 & $0.13 \pm 0.05$ & $-0.33 \pm 0.04$ \\
& \multicolumn{2}{l}{5. Intersection of PA45, PA331}\\
PA45  & $0.44 \pm 0.11$ & $-0.79 \pm 0.04$ \\
PA331 & $0.48 \pm 0.04$ & $-0.85 \pm 0.03$ \\
& \multicolumn{2}{l}{6. Intersection of PA45, PA132}\\
PA45  & $0.61 \pm 0.21$ & $-1.06 \pm 0.11$ \\
PA132 & $0.48 \pm 0.07$ & $-0.81 \pm 0.23$ \\
\hline
\end{tabular}
\end{center}
\end{table*}

\begin{table*}[t!]

\caption{Emission line flux ratios} \medskip

\begin{tabular}{llccccc}
\hline
\multicolumn{1}{l}{Region}&\multicolumn{1}{l}{Slit/position}   &
$\frac{I([\textrm{OIII}]5007)}{I(\textrm{H}\beta)}$ &
$\frac{I([\textrm{SII}]6717+6731)}{I(\textrm{H}\alpha)}$ &
$\frac{I([\textrm{NII}]6583)}{I(\textrm{H}\alpha)}$ &
$\frac{I([\textrm{SII}]6717}{I([\textrm{SII}]6731}$ &
$\frac{I([\textrm{OII}]3727+3729)}{I([\textrm{OIII}]5007)}$ \\
\hline
HL111a& PA0 [$-13$, $-9$] & $1.90 \pm 0.03$ & $0.178 \pm 0.002$ & $0.134 \pm 0.003$ & $1.39 \pm0.03$ & $2.64 \pm0.33$    \\
      & PA331 [4, 11] & $2.09 \pm 0.12$ & $0.201 \pm 0.003$ & $0.099 \pm 0.003$     & $1.10 \pm0.01$ & $1.44 \pm0.22$    \\
HL111b& PA0 [$-20$, $-15$]& $1.51 \pm 0.05$ & $0.183 \pm 0.03$  & $0.148 \pm 0.003$ & $1.47 \pm0.05$ & $3.63 \pm0.41$    \\
HL111c& PA0 [$-4$, $3$]   & $3.52 \pm 0.02$ & $0.081 \pm 0.001$ & $0.069 \pm 0.001$ & $1.41 \pm0.03$ & $0.80 \pm0.10$    \\
      & PA268 [$-6$, $5$] & $3.47 \pm 0.01$ & $0.063 \pm 0.001$ & $0.047 \pm 0.001$ & $1.40 \pm0.01$ & $0.50 \pm0.02$    \\
HL111d& PA268 [$-12$, $-8$]& $2.95 \pm 0.01$ & $0.137 \pm 0.001$ & $0.097 \pm 0.001$& $1.41 \pm0.02$ & $1.16 \pm0.08$    \\
HL111e& PA0 [$-7$, $-5$]  & $2.44 \pm 0.06$ & $0.138 \pm 0.002$ & $0.101 \pm 0.002$ & $1.50 \pm0.05$ & $1.62 \pm0.41$    \\
HL106& PA0 [$-42$, $-21$] & $2.89 \pm 0.07$ & $0.161 \pm 0.004$ & $0.117 \pm 0.004$ & $1.46 \pm0.08$ & $1.40 \pm0.41$    \\
     & PA132 [62, 80] & $2.31 \pm 0.03$ & $0.191 \pm 0.007$ & $0.102 \pm 0.006$     & $1.26 \pm0.09$ & $1.23 \pm0.08$    \\
     & PA331 [$-20$, $-8$]& $2.42 \pm 0.11$ & $0.144 \pm 0.002$ & $0.066 \pm 0.02$  & $1.02 \pm0.01$ & $1.11 \pm0.41$    \\
SS& PA0 [$-60$, $-55$]    & $0.75 \pm 0.10$ & $0.713 \pm 0.017$ & $0.247 \pm 0.013$ & $1.58 \pm0.08$ & $4.24 \pm0.44$    \\
  & PA132 [84, 100]   & $1.33 \pm 0.13$ & $0.745 \pm 0.100$ & $ 0.230 \pm 0.056$    & $1.24 \pm0.26$ & $3.49 \pm0.42$    \\
  & PA331 [$-40$, $-27$]  & $1.87 \pm 0.10$ & $0.625 \pm 0.006$ & $0.207 \pm 0.04$  & $1.63 \pm0.09$ & $1.99 \pm0.43$    \\
HL37& PA45 [$-12$, $-6$]  & $3.34 \pm 0.24$ & $0.073 \pm 0.002$ & $0.033 \pm 0.012$ & $1.55 \pm0.05$ & $0.87 \pm0.26$    \\
    & PA132 [$-52$, $-44$]& $2.85 \pm 0.05$ & $0.180 \pm 0.030$ & $0.064 \pm 0.024$ & $0.99 \pm0.22$ & $0.89 \pm0.13$    \\
HL45& PA45 [$-4$, $3$]    & $4.86 \pm 0.09$ & $0.079 \pm 0.005$ & $0.045 \pm 0.002$ & $1.12 \pm0.03$ & $0.22 \pm0.05$    \\
HL50& PA45 [12, 27]   & $3.44 \pm 0.08$ & $0.073 \pm 0.005$ & $0.045 \pm 0.002$     & $1.35 \pm0.05$ & $0.69 \pm0.14$    \\
    & PA331 [108, 115]& $3.05 \pm 0.08$ & $0.162 \pm 0.009$ & $0.086 \pm 0.002$     & $1.49 \pm0.14$ & $1.11 \pm0.10$    \\
HL100& PA132 [49, 58] & $1.65 \pm 0.02$ & $0.364 \pm 0.020$ & $0.179 \pm 0.023$     & $1.24 \pm0.11$ & $2.69 \pm0.23$    \\
HL89& PA132 [35, 43]  & $1.32 \pm 0.2$ & $0.446 \pm 0.029$ & $0.168 \pm 0.039$      & $1.19 \pm0.12$ & $4.39 \pm0.32$    \\
 M10& PA132 [20, 30]& $1.49 \pm 0.06$ & $0.753 \pm 0.065$ & $0.165 \pm 0.026$       & $1.03 \pm0.13$ & $6.41 \pm0.53$    \\
HL67& PA132 [5, 15]   & $1.61 \pm 0.06$ & $0.578 \pm 0.076$ & $0.142 \pm 0.044$     & $0.89 \pm0.15$ & $4.13 \pm0.48$    \\
HL41& PA132 [$-43$, $-39$]& $1.58 \pm 0.05$ & $0.254 \pm 0.011$ & $0.126 \pm 0.013$ & $1.07 \pm0.09$ & $0.43 \pm0.41$    \\
HL36& PA132 [$-61$, $-55$]& $2.86 \pm 0.03$ & $0.137 \pm 0.019$ & $0.065 \pm 0.007$ & $0.94 \pm0.22$ & $0.60 \pm0.17$    \\
HL22& PA132 [$-116$, $-108$]& $1.53 \pm 0.03$ &$0.158\pm 0.011$ & $0.126 \pm 0.007$ & $1.17 \pm0.13$ & $2.60 \pm0.37$    \\
HL46\_48&PA268 [68, 71]& $1.49 \pm 0.08$ & $0.320 \pm 0.011$ & $ 0.200 \pm 0.010$    & $1.66 \pm0.09$ & $2.94 \pm0.43$    \\
\hline
\end{tabular}
\end{table*}

\begin{table*}[t!]

\caption{Relative O, N$^+$, and S$^+$ abundances} \medskip

\begin{tabular}{llcccc}
\hline
\multicolumn{1}{l}{Region}& \multicolumn{1}{l}{Slit/position } &
$12+\log(\textrm{O/H})$ (1) & $12+\log(\textrm{O/H})$ (2) &
$12+\log(\textrm{N}^{+}/\textrm{H})$ &
$12+\log(\textrm{S}^{+}/\textrm{H})$\\
\hline
HL111a& PA0 [$-13$, $-9$] & $8.34 \pm 0.24$ & $8.28 \pm 0.12$ & $6.26 \pm 0.13$ & $5.46 \pm 0.12$ \\
      & PA331 [4, 11] & $8.05 \pm 0.34$ & $8.15 \pm 0.15$ & $6.54 \pm 0.16$ & $6.00 \pm 0.16$ \\
HL111b& PA0 [$-20$, $-15$]& $8.27 \pm 0.22$ & $8.29 \pm 0.13$ & $6.38 \pm 0.13$ & $5.55 \pm 0.13$ \\
HL111c& PA0 [$-4$, $3$]   & $8.24 \pm 0.15$ & $8.24 \pm 0.10$ & $5.92 \pm 0.10$ & $5.07 \pm 0.10$ \\
      & PA268 [$-6$, $5$] & $7.94 \pm 0.10$ & $8.15 \pm 0.10$ & $6.05 \pm 0.12$ &  $6.56 \pm 0.22$\\
HL111d&PA268 [$-12$, $-8$]& $8.28 \pm 0.11$ & $8.05 \pm 0.16$ & $6.10 \pm 0.18$ & $5.33 \pm 0.19$ \\
HL111e& PA0 [4--7, $-5$]  & $8.29 \pm 0.38$ & $8.24 \pm 0.18$ & $6.21 \pm 0.19$ & $5.42 \pm 0.18$ \\
HL106& PA0 [$-42$, $-21$] & $8.41 \pm 0.42$ & $8.28 \pm 0.22$ & $6.21 \pm 0.22$ & $5.41 \pm 0.22$ \\
     & PA132 [62, 80] & $8.06 \pm 0.14$ & $8.15 \pm 0.10$ & $6.01 \pm 0.10$ &  $5.32 \pm 0.10$ \\
     & PA331 [$-20$, $-8$]& $8.03 \pm 0.42$ & $8.14 \pm 0.27$ & $6.39 \pm 0.27$ & $5.88 \pm 0.28$ \\
SS& PA0 [$-60$, $-55$]    & $7.59 \pm 0.42$ & $8.04 \pm 0.43$ & $6.44 \pm 0.41$ & $5.97 \pm 0.43$ \\
  & PA132 [84, 100]   & $8.07 \pm 0.18$ & $8.22 \pm 0.21$ & $6.45 \pm 0.38$ & $6.00 \pm 0.41$ \\
  & PA331 [$-40$, $-27$]  & $8.12 \pm 0.42$ & $8.19 \pm 0.39$ & $6.88 \pm 0.39$ & $6.50 \pm 0.39$ \\
HL37& PA45 [$-12$, $-6$]  & $8.23 \pm 0.42$ & $8.23 \pm 0.20$ & $6.23 \pm 0.24$ & $5.63 \pm 0.21$ \\
    & PA132 [$-52$, $-44$]& $8.07 \pm 0.23$ & $8.17 \pm 0.10$ & $5.96 \pm 0.22$ & $5.39 \pm 0.17$ \\
HL45& PA45 [$-4$, $3$]    & $7.92 \pm 0.17$ & $8.21 \pm 0.10$ & $6.24 \pm 0.12$ & $5.60 \pm 0.10$ \\
HL50& PA45 [12, 27]   & $8.10 \pm 0.36$ & $8.20 \pm 0.13$ & $6.32 \pm 0.14$ & $5.58 \pm 0.13$ \\
    & PA331 [108, 115]& $8.30 \pm 0.16$ & $8.25 \pm 0.10$ & $6.56 \pm 0.10$ & $5.80 \pm 0.10$ \\
HL100& PA132 [49, 58] & $8.16 \pm 0.16$ & $8.23 \pm 0.10$ & $6.31 \pm 0.15$ & $5.65 \pm 0.10$ \\
HL89& PA132 [35, 43]  & $8.23 \pm 0.13$ & $8.29 \pm 0.10$ & $6.30 \pm 0.27$ &  $5.72 \pm 0.10$\\
 M10& PA132 [20, 30]& $8.74 \pm 0.36$ & $8.49 \pm 0.19$ & $6.43 \pm 0.29$ &  $6.04 \pm 0.20$\\
HL67& PA132 [5, 15]   & $8.52 \pm 0.23$ & $8.36 \pm 0.13$ & $6.34 \pm 0.20$ & $5.86 \pm 0.17$ \\
HL41& PA132 [$-43$, $-39$]& $7.73 \pm 0.42$ & $7.79 \pm 0.39$ & $6.24 \pm 0.39$  & $5.55 \pm 0.39$ \\
HL36& PA132 [$-61$, $-55$]& $7.86 \pm 0.42$ & $8.09 \pm 0.15$ & $5.96 \pm 0.19$ & $5.28 \pm 0.19$ \\
HL22& PA132 [$-116$, $-108$]& $8.28 \pm 0.12$ & $8.18 \pm 0.14$ & $6.25 \pm 0.16$ &  $5.44 \pm 0.16$\\
HL46\_48&PA268 [68, 71]& $8.10 \pm 0.40$ & $8.21 \pm 0.21$ & $6.58 \pm 0.23$ & $5.85 \pm 0.21$ \\
\multicolumn{6}{c}{from observations by Lequeux et~al.~(1979)}\\
HL111&   & 7.74 & 8.17 &  6.23 & 5.48 \\
HL45&   & 7.92 & 8.31 &  6.24 & 5.40\\
\hline
\end{tabular}
\end{table*}

For the abundance analysis, we employ equation (24) from Pilyugin and
Thuan~(2005), which is the revisited calibration dependence of the
oxygen abundance $12 + \log(\textrm{O/H})$ on the flux ratios of
strongest [OII] and [OIII] lines. The
classical method of estimation from the so-called oxygen
abundance indicator~$R_{23}$ suggested by Pagel et~al.~(1979) and
widely used previously was modified in a series of papers by
Pilyugin (see Pilyugin and Thuan~(2005)) and is based on two
parameters: $R_{23}$ and the excitation parameter~$P$.
For verification purposes, we also incorporate other techniques for the
abundance estimation developed by Izotov et al.(2006) and by Charlot and
Longhetti (2001).

The distributions of abundances $12+ \log(\textrm{O/H})$ along
the five long slits obtained are shown in Fig.~5a and presented in
Table~4. The mean oxygen abundances in the bright HII regions are
shown by dots with the error bars.

As was noted by Pilyugin and Thuan~(2005) their (24) relation is
most reliable for~$P$ in the range from $\simeq 0.55$ to $\simeq
1$. Having eliminated the points corresponding to $P<0.55$, we
verified that this did not change significantly our averaged
estimates of the oxygen abundance but only slightly increased the
measurement accuracy.

To estimate the relative nitrogen and sulfur abundances in
various HII~regions, we used Eqs.~(6) and~(8) from Izotov
et~al.~(2006)

The resulting sulfur and nitrogen abundances in our regions of IC~10
are shown in Fig.5a and 5b, respectively.
Note that we actually estimate the S$^{+}$ abundance,
because we disregarded the S$^{2+}$ emission in the
$[\textrm{SIII}]\lambda6312$ line that we can measure only in~HL111c.
This is also true for the nitrogen abundance.

To properly estimate the N/O and S/O ratios, we also found the
relative oxygen abundance $12 + \log(\textrm{O/H})$ using Eqs.~(3)
and~(5) from Izotov et~al.~(2006)

The characteristic electron density in galactic ionized-gas regions that we
found from the [SII] line flux ratio varies within the range from~30 to
300--400~cm$^{-3}$ and these variations have virtually no effect on the
abundance estimates.

We assume the electron temperature to be
$T_e(\textrm{OIII})=11000$~K in all HII~regions.
There are three reasons for choosing this value.
First, it is close to a temperature of about 11600~K that we determined from
the oxygen line flux ratio in~HL111c. This is the only bright HII region
where we can detect the [OIII]$\lambda4363$~\AA\ auroral line.
Unfortunately, the line flux has a large uncertainty, because the
line is weak. Second, the adopted value lies between
the ones obtained by Lequeux et~al.~(2003) and Lee et~al.~(2003) for the
three brightest HII~regions in the galaxy. Third, this electron temperature
provides
the best agreement between the oxygen abundances derived
according to Pilyugin and Thuan~(2005), and Izotov et~al.(2006)
The temperature in the [OII] emission region was found
from the relation (Izotov et~al. 2006)

$$
t(\textrm{OII})=-0.744+t(\textrm{OIII})\cdot(2.338-0.610t(\textrm{OIII})).
$$

We assume the same temperature for the [SII] and [NII] emission region.

The obtained results are collected in Table~4. Its columns show
the designation of the HII~region according to the catalog
by Hodge and Lee~(1990), the spectrogram used for our estimates
and the positions on it, and the derived mean oxygen,
nitrogen, and sulfur abundances. The oxygen abundances estimated
using Eq.~(24) from Pilyugin and Thuan~(2005) are denoted in Table~4
by (1). Those estimated with Eqs.~(3) and~(5) from
Izotov et~al.~(2006) are denoted by (2).

We did not average the abundances found from different spectra for the same
HII~region, since they correspond to different slit positions on the nebula
and could differ due to different physical conditions: the density and porosity
of interstellar medium, loci of the radiation sources, etc.

The lower rows in Table~4 show the abundances in two regions:
HL~111 and HL~45 which correspond to the regions 1 and 2, from
Fig.~1 by Lequeux et~al. (1979). We derived the abundances for the
relative line intensities measured by Lequeux et~al.~(1979). In
both cases the slit in this paper was in the east--west direction.
The region over which the relative abundances were averaged spans
3".8 $\times$ 12".4. Since no localization of the spectrograms is
given by Lequeux et~al.(1979), we can make the comparison only
approximately.

\section*{DISCUSSION}

\subsection*{Interstellar Extinction}

Variations of the color excess that we revealed here for
different galactic regions lie within the range 0.8 - 1.1 mag.
This is considerably smaller than the difference from literature,
where it ranges from $E(B - V)$ = 0.47 mag to $E(B - V)$ = 2.0
mag (see Sakai et~al. 1999; Demers et~al.~2004; and references
therein). In the three brightest HII~regions, Lee et~al.~(2003)
found $E(B - V)$ = 0.79  -- 1.2 mag, which is similar to our
estimates. Variations of $E(B - V)$ that we observed can be
naturally explained by two factors. First, a local higher
extinction is possible, since the brightest star-forming region
is projected to the densest cloud of neutral hydrogen and
molecular gas in the galaxy, and hence can be partly embedded
into it. A similar effect was also observed by other authors. In
particular, the extinction estimated by Borissova et~al.~(2000)
from the flux ratio of the Br$\gamma$ and H$\alpha$ lines in the
two brightest nebulae~HL111 and~HL45 is considerably higher than
that found from stars. Second, a local lower extinction is
possible if the stellar wind and SN explosions in young groups
sweep out the surrounding dense gas. In particular, in the
central star-forming region Vacca et~al.~(2007) found $E(B - V)$
= 0.6 mag from massive blue stars and concluded that the young
stars born in the last star formation episode are located at the
front boundary of the above-mentioned dense cloud in front of the
old stellar population.

In this paper we present one of the most detailed studies of the
oxygen, nitrogen, and sulfur abundances in IC~10. Previously,
Lequeux et~al.~(1979) found $12+\log(\textrm{O/H})=8.17$
for~HL111 and $12+\log(\textrm{O/H})=8.45$ for~HL45. The lower
rows in Table~4 give our estimates based on the measurements by
Lequeux et~al.~(1979). According to Garnett~(1990),
$12+\log(\textrm{O/H})=8.19$ for~HL45. According to Richer
et~al.~(2001), $12+\lg(\textrm{O/H})=8.23\pm 0.09$ in the
brightest nebula HL111c. For the brightest objects with detected
[OIII]$\lambda4363$~\AA\ line, Lee et~al.~(2003) found that
$12+\log(\textrm{O/H})=8.32\pm 0.14$ in HL111c, $8.05\pm 0.18$ in
HL111e, $12+\log(\textrm{O/H}) \geq 7.96$ in HL111b, and
$12+\log(\textrm{O/H}) \geq 7.60$ in the clump~HL106a
inside~HL106. In the latter two objects, only the upper abundance
limit was estimated. As we see, all listed results are in a good
agreement with our observations of HII regions in IC~10.

As our measurements showed, the relative oxygen line intensities and,
accordingly, the relative oxygen abundances vary
in a wide range, from~7.6 to~8.5. To understand whether
it is related to the observational errors or to the actual differences
between HII~regions, we presented the relative intensities found, respectively,
for bright and faint objects in Figs.~4a and~4b. As follows from the figure,
the measurement errors are slightly larger for the faint regions than
that in the bright ones. However, the scatter of values in the faint
regions does not
exceed significantly that for the bright ones.
Therefore, we conclude that the oxygen abundance
variations in the galaxy are most likely real. At the same time, the systematic
difference between the relative intensities in the bright and faint regions is
observed (see below).

The relative oxygen abundance averaged over all galactic regions that we
investigated here is $12+\log(\textrm{O/H}) = 8.17 \pm 0.35$, which corresponds
to the metallicity $Z=0.18 \pm 0.14 Z_{\odot}$.

\begin{figure*}[t!]
\begin{center}
\includegraphics[scale=1.0]{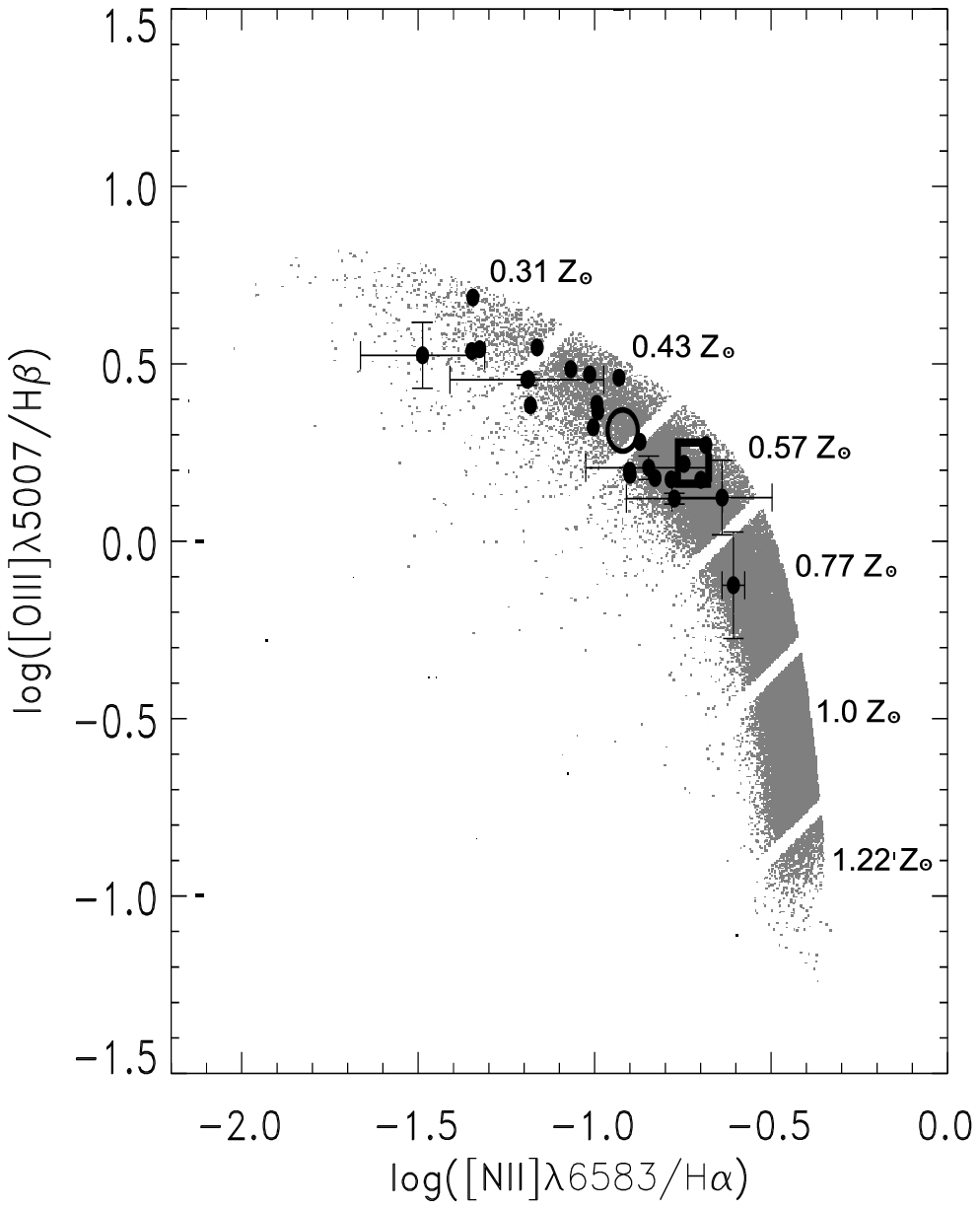}
\end{center}

\caption{Comparison of our line intensity ratios with results by
Cid~Fernandes
et~al.~(2007) and Asari et~al.~(2007). The mean metallicity in each of the
gray sectors is shown to the right of the localization band of
85000~starburst~galaxies. The circle and the square indicate the
galaxy-averaged values found by Hidalgo-Gamez~(2005) for bright HII~regions and
the diffuse ionized gas~(DIG), respectively. \hfill}
\end{figure*}

\subsection*{Diagnostic Diagrams of Relative Line Intensities}

The relative intensities of diagnostic lines in the bright nebulae and faint
regions of diffuse ionized gas (DIG) in the whole
galaxy and over its four
fields taken apart were estimated previously by Hidalgo-Gamez~(2005) from nine spectrograms
combined into one wide band. According to this paper, the galaxy-averaged
relative intensities are $I([OIII]5007\textrm{\AA}) / I(\textrm{H}\beta)=1.86$,
$I$($[\textrm{NII}]\lambda6583$\AA) / $I(\textrm{H}\alpha)=0.16$, and
$I$($[\textrm{SII}]\lambda6717$\AA) / $I$($\textrm{H}\alpha)=0.28$. The values
averaged over the bright nebulae are
$I$($[\textrm{OIII}]\lambda5007$\AA) / $I(\textrm{H}\beta)=2.05$,
$I$($[\textrm{NII}]\lambda6583$\AA) / $I(\textrm{H}\alpha)=0.12$,
$I$($[\textrm{SII}]\lambda6717$\AA) / $I(\textrm{H}\alpha)=0.19$.
The values averaged over the faint diffuse ionized gas (DIG) are
$I$($[\textrm{OIII}]\lambda5007$\AA) / $I(\textrm{H}\beta)=1.66$,
$I$($[\textrm{NII}]\lambda6583$\AA) / $I(\textrm{H}\alpha)=0.19$,
$I$($[\textrm{SII}]\lambda6717$\AA) / $I(\textrm{H}\alpha)=0.31$.
Our measurements of these line ratios in individual HII~regions
are in good agreement with the values above, given the
observational errors.

As follows from Figs.~4a and~4b, similar to other Irr~galaxies investigated in
detail, the diagnostic diagrams slightly differ between the bright and
faint regions
of~IC~10: the relative intensities of [SII] lines are appreciably higher in
faint regions as well as those of [NII] lines are also slightly higher,
whereas
$I$([OIII]$\lambda5007$~\AA)/$I$(H$_\beta$) weakens in these regions.
The similar picture is
also observed in spiral galaxies, for example, in~M31 (see Galarza et~al.
1999). The value of $I$([SII])/$I$(H$\alpha$) in the
bright compact HII~regions is
lower than that in the faint diffuse and ring nebulae, not to mention the DIG.
The
noted differences can be explained by a decrease in the ionization parameter
when transiting from the compact to faint diffuse regions.
This effect is predicted by
photoionization models. The contribution from the gas emission behind the front
of the shocks triggered by supernova explosions and stellar wind also leads to
the same effect of the relative intensity
$I$([SII])/$I$(H$\alpha$) enhancement. The filamentary structure of the entire H$\alpha$
emission region in the galaxy suggests that the shocks play a prominent
role in IC~10.

It should be noted that the constructed diagnostic lines diagrams
agree poorly with the currently available photoionization models
for the metallicity $Z=0.2 Z_{\odot}$ found above from strong
oxygen lines. In particular, the observed diagnostic diagrams are
consistent with the families of theoretical diagnostic curves
(Dopita et~al. 2006) only for a metallicity from $Z=0.4
Z_{\odot}$ to $Z$= (1 - 2) $Z_{\odot}$.

In Fig.~6, our data are compared with the results of Cid~Fernandes
et~al.~(2007) and Asari et~al.~(2007), who summarized the gas metallicity
estimates for $\simeq85000$ starburst~galaxies. On the
$I([\textrm{OIII}]\lambda 5007\textrm{\AA}) / I(H\beta)$ versus
$I([\textrm{NII}]\lambda6583\textrm{\AA}) / I(\textrm{H}\alpha)$ diagram, these
authors give the number of objects in each sector of the galaxy localization
band and the mean metallicity that corresponds to the sector. As we see, the
HII~regions in IC~10 fall nicely on the galaxy localization band but fall
within the range of metallicities from $Z=0.3 Z_{\odot}$ to $Z= 0.6 Z_{\odot}$.
One distinctive region in IC~10, the synchrotron superbubble, falls
near $Z= 0.8 Z_{\odot}$.
This object is discussed below. Note, however, that Asari et~al.~(2007)
recorded a systematic shift by 0.2~dex between their metallicity estimates and
those of Pilyugin and Thuan~(2005) and Izotov et~al.~(2006) for 177~``common''
objects made from the relations that we used here. It turns out that the lower
the metallicity, the larger the shift. For this reason, Asari et~al.
adopted the
metallicity $Z=0.2 Z_{\odot}$ for Sector~A rather than $Z=0.3 Z_{\odot}$ and
therefore $Z=0.2 Z_{\odot}$ is the lower boundary of the range into which
our regions of IC~10 fall.

It is also interesting to compare the observed diagnostic diagrams with the
theoretical photoionization models by Charlot and Longhetti~(2001), since these
authors separately took into account the changes in each of the parameters
defining the model calculations, such as the cluster age, the initial mass
function (IMF), ionization parameter, density of interstellar medium, and
metallicity.

\begin{figure*}[p!]
\begin{center}
\includegraphics[scale=0.65]{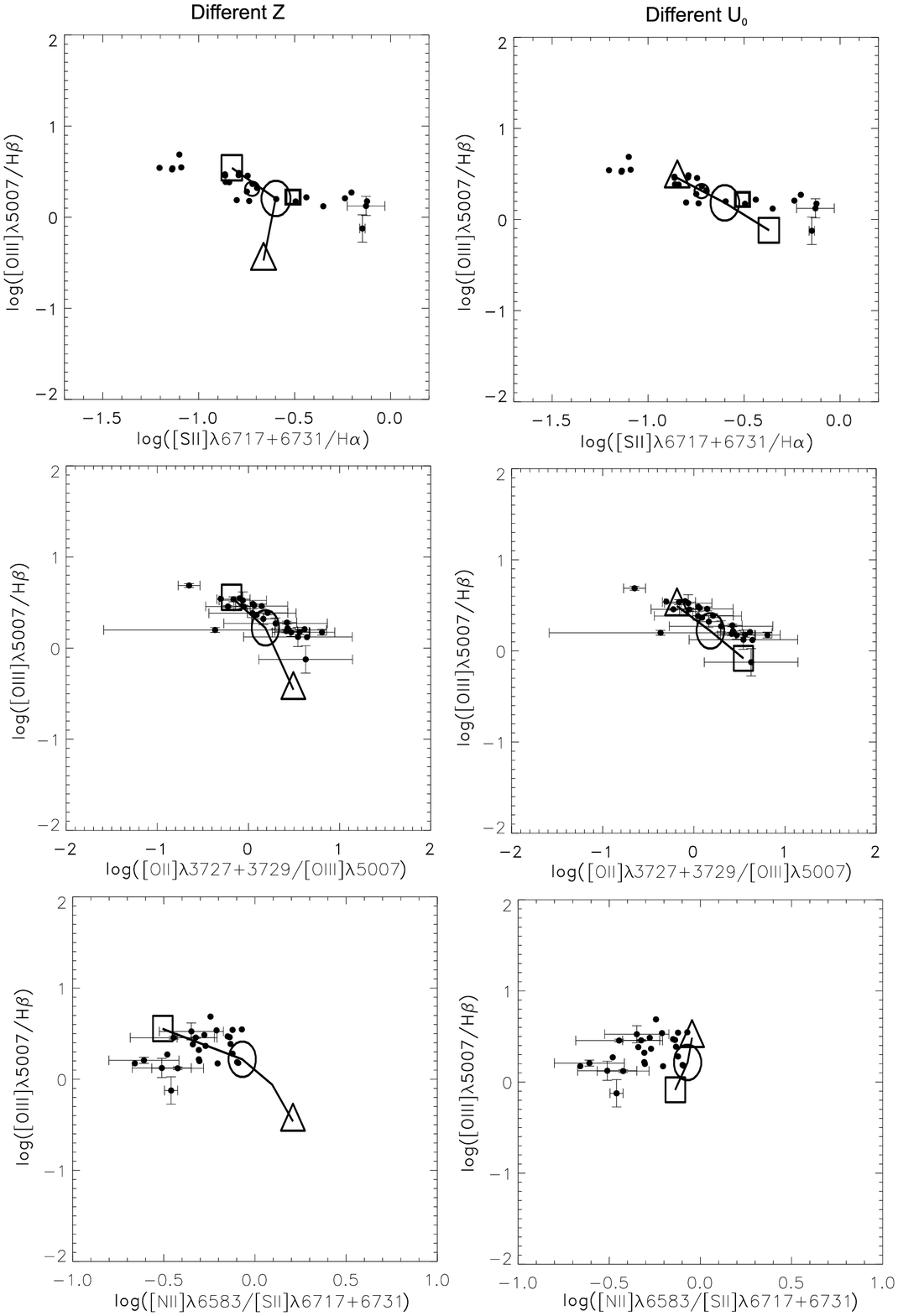}
\end{center}
\caption{Comparison of the line flux ratios on diagnostic
diagrams with the model calculations by Charlot and
Longhetti~(2001). The large squares, circles, and triangles in the
left panels~(a) mark the regions of reduced ($Z=0.2 Z_{\odot}$),
solar ($Z= Z_{\odot}$), and enhanced ($Z=2 Z_{\odot}$)
metallicity, respectively, in accordance with the model
calculations. The same symbols in the right panels~(b) indicate
the model calculations for the effective ionization parameters
$\log U_{0} -3.0$, $-2.5$, and $-2.0$, respectively. \hfill}
\end{figure*}

Comparison of the diagnostic diagrams for~IC~10 with the distributions of
${I([\textrm{OIII}]/I(\textrm{H}\beta)}$ as a function of
$I$([OII])/$I$([OIII]), $I$([SII])/$I$(H$\alpha$), and $I$([NII])/$I$([SII])
constructed by Charlot and Longhetti~(2001) for galaxies of various types and
HII~regions showed an excellent agreement between them and our observing
line flux ratios.
In Fig.~7 our observations are compared with the models computed by Charlot and
Longhetti~(2001) for various metallicities and various effective ionization
parameters. The large square, circle, and triangle on our diagnostic diagrams
(left panels) mark the regions of reduced ($Z=0.2 Z_{\odot}$), solar ($Z=
Z_{\odot}$), and enhanced ($Z=2 Z_{\odot}$) metallicity from models
by Charlot and Longhetti~(2001), see Figs.~1, 2, and~3 in their
paper. The large square, circle, and triangle on the right panels indicate
three values of the model flux ratios for the effective ionization
parameter $\log U_{0}$ of $-3.0$,
$-2.5$, and $-2.0$ from the calculations by Charlot and
Longhetti~(2001). As we see, in complete agreement with the model calculations,
the $I([\textrm{OIII}]\lambda 5007\textrm{\AA})/I(\textrm{H}\beta)$ versus
$I([\textrm{NII}]\lambda6583\textrm{\AA})/I([\textrm{SII}]\lambda6717+6731\textrm{\AA})$
diagnostic diagrams are most promising for estimating the metallicity of the
galactic gaseous medium.
Positions of our HII~regions in~IC~10 on
the diagnostic diagrams
correspond to the metallicities within the range of $Z$= (0.2 -- 1)
$Z_{\odot}$. The dependences of $I([\textrm{OIII}]\lambda
5007\textrm{\AA})/I(\textrm{H}\beta)$ on $I$([OII])/$I$([OIII]) and of
$I([\textrm{OIII}]\lambda 5007\textrm{\AA})/I(\textrm{H}\beta)$ on
$I([\textrm{SII}]\lambda6717+\linebreak+6731\textrm{\AA})/I(\textrm{H}\alpha)$
are well explained by variations in the ionization parameter.

The differences between the metallicity of individual HII~regions in the galaxy
found from the $R_{23}$ parameter and the metallicity
determined by comparing diagnostic diagrams with
theoretical photoionization models can be explained by a number of factors.
First of all, the photoionization models are computed for rich star clusters.
In particular, Dopita et~al.~(2006) assumed a cluster mass of $M \geq 10^{3}
M_{\odot}$,  the IMF by Miller and Scalo ranges from~1 to $100M_{\odot}$, while
the star clusters in~IC~10 are comparatively poor (see Hunter~2001).

In addition, the contribution from the gas emission behind the front of the
shocks can be significant in IC~10, as suggested by the filamentary structure
of the entire star-forming region in this galaxy. Actually, the term ``diffuse
ionized gas''~(DIG) can be applied to IC~10 only conditionally, because the
faint emission regions are characterized mostly as
filamentary shell and/or arc structure. Therefore, the faintest regions inside
a number of shells that appear diffuse most likely represent the front or back
sides of the shells.

The synchrotron superbubble in IC~10 is a unique object and deserves a
more detailed discussion. According to Lozinskaya and Moiseev~(2007) and
Lozinskaya et~al.~(2008), the image in~[SII] lines better than in H$\alpha$
reveals the optical shell
identified with the synchrotron superbubble. The
expansion velocity of the bright knots and filaments in the optical shell found
in these papers, 50--80~km~s$^{-1}$, confirms the estimate by Ramsey
et~al.~(2006). The gas mass is $M\simeq 4\times 10^{5} M_{\odot}$ and the
kinetic energy is $E_{\textrm{kin}}\simeq (1-3)\cdot 10^{52}$~erg.
This amount of released
energy corresponds to the explosions of ten supernovae plus the stellar wind
from their host association. Alternatively, the superbubble may be a result
of a hypernova
explosion, as was suggested by Lozinskaya and Moiseev~(2007). The kinematic age
of the superbubble, $t \simeq$ (3 -- 7)$\cdot 10^{5}$~yr, is a strong
argument for a hypernova, because it takes more than $t \simeq 10^{7}$~yr for
the explosions of ten supernovae in a local galactic region.

As it can be seen in Fig.~1, the regions from~$85''$ to~$120''$ on spectrogram~PA132,
from~$-60''$ to~$-25''$ on spectrogram~PA331, and from~$-50''$ to~$-80''$ on
spectrogram~PA0 (the latter passes over the very edge of the superbubble)
correspond to the synchrotron superbubble.

Our measurements indicate that the relative intensity
$I([\textrm{SII}]\lambda6717+6731\textrm{\AA})/I(\textrm{H}\alpha)$ in the
region of the synchrotron superbubble reaches 0.7--0.75,
which strongly suggests a
gas emission behind the shock front (Dopita and Sutherland~1995,~1996). This is
also suggested by the ratios of $I([\textrm{NII}]\lambda
6583\textrm{\AA})/I(\textrm{H}\alpha) \simeq 0.25$ and
$I([\textrm{OII}]\lambda3727+3729\textrm{\AA})/I([\textrm{OIII}]\lambda5007\textrm{\AA})\simeq
2 -  4$ that are higher than
those in the nearby bright region~HL106, and also by the lower
relative intensity
$I([\textrm{OIII}]\lambda5007\textrm{\AA})/I(\textrm{H}\beta)$.

The region of the synchrotron superbubble on the diagnostic
diagrams also differs from HII~regions by a lower ratio
$I([\textrm{OIII}]\lambda5007\textrm{\AA}) / I(\textrm{H}\beta)$
and by high $I([\textrm{SII}]\lambda6717+6731\textrm{\AA}) /
I(\textrm{H}\alpha)$, \\
$I([\textrm{NII}]\lambda6583\textrm{\AA}) / I(\textrm{H}\alpha)$,
and \linebreak $I([\textrm{OII}]\lambda3727+3729\textrm{\AA}) /
I([\textrm{OIII}]\lambda5007\textrm{\AA})$. This suggests the
dominance of the collisional excitation and is typical in old
supernova remnants.

Lozinskaya and Moiseev~(2007) found a mean $n_{e}\simeq $ 20 --
30~cm$^{-3}$ from the intensities of the
[SII]($\lambda6717/6731$\AA) lines
for the brighter northern part of the superbubble
(the region 90$''$ -- 105$''$ on~PA132), where the errors are
relatively small. For the region $-50''$ --
$-60''$ on spectrogram~PA0, we find a similar mean density $n_{e}
\simeq$ 30 -- 40~cm$^{-3}$. In the southern faint part of the
superbubble, individual knots can be distinguished in the
spectrogram~PA132 in which the density reaches $n_{e} \simeq$ 200
-- 300~cm$^{-3}$, but the accuracy of these estimates is low. The
gas density northward of the synchrotron superbubble (the
positions range between $80''$ --  90$''$ on~PA132) is $n_{e} \simeq
200$~cm$^{-3}$.

The oxygen abundance estimate in the region of the synchrotron superbubble is
ambiguous. In general, the present-day models of the gas emission spectrum
behind the shock front developed by Allen et~al.~(2008) allow the abundances of
heavy elements, in particular oxygen, to be determined from the available
diagnostic relative line intensities if the ambient gas density and the shock
velocity are known. However, in our case the problem is complicated by the
presence of the WR star~M17 in the central region of the synchrotron
superbubble, whose ionizing radiation may be significant or even
dominant in close neighborhoods. On the other hand, we cannot properly use the
technique applied above to HII~regions without taking into account the gas
emission behind the shock front.

A careful calculation including both emission mechanisms is
beyond the scope of this paper. Below we formally estimate the
oxygen abundance in the synchrotron superbubble independently by
two methods, realizing that both are improper. Employing the
$R_{23}$ parameter, we found the abundances
$12+\log(\textrm{O/H})=8.07\pm 0.18$ from spectrogram~PA132,
$7.59\pm 0.42$ from~PA0, and $8.12\pm 0.42$ from~PA331 using the
technique of Pilyugin and Thuan~(2005) or, respectively, $8.22\pm
0.21$ from spectrogram~PA132, $8.04\pm 0.43$ from~PA0, and
$8.19\pm 0.39$ from~PA331 using the technique of Izotov
et~al.~(2006) (see Table~4). Since the measurements were made for
three different regions of a very faint extended object, these
independent estimates of the oxygen abundance are in a very good
agreement.

A detailed kinematic study of the synchrotron superbubble (Lozinskaya
et~al.~2008) clearly revealed weak H$\alpha$ wings in the velocity range
from~$-200$ to $-450$~km~s$^{-1}$. These high-velocity features in the line
correspond to the velocity of the shock wave triggered by a hypernova explosion of
about 100--200~km~s$^{-1}$. Comparison with the models of gas emission behind
the front of a shock (Allen et~al.~2008) propagating with such velocity gives
the best agreement at the abundance $12+\log(\textrm{O/H}) \simeq 8.15$ with
the models of purely collisional excitation,
or $12+\log(\textrm{O/H}) \simeq 8.35$
with the models that include the gas pre-ionization by a shock.

\section*{CONCLUSIONS}

We present the most detailed to-date study of
the O, N$^+$, and S$^+$ abundances of the
individual HII~regions and the synchrotron superbubble in the
dwarf~Irr~galaxy IC~10
that we estimated from five spectrograms taken with the SCORPIO instrument
at the 6-m SAO telescope.

The relative~O, N$^+$, and S$^+$
abundances in individual HII~regions of the galaxy IC~10 lie
within the following ranges: $12 + \log(\textrm{O/H})$ = 7.59 --
8.52 , $12 + \log(\textrm{N}^{+}/\textrm{H})$ = 5.92 -- 6.58 ,
and $12 + \log(\textrm{S}^{+}/\textrm{H})$ = 5.00 --  6.00. The
mean metallicity of the galaxy's gaseous medium is $Z = 0.18
\pm 0.14 Z_{\odot}$. The metallicity estimates from the comparison
of diagnostic diagrams with photoionization models are found to
be less reliable than those from the strong oxygen line ratios.

The oxygen abundance estimates in the synchrotron superbubble
from the measured relative line intensities must be based on the
model of the combined action of a shock (the result of a
hypernova explosion) and ionizing radiation (the presence of a WR
star in the central region). So far the estimates have been made
separately within the framework of these two models. With the model
of an HII~region, we found that $12+\log(\textrm{O/H})=$7.59 -- 8.34.
The comparison with the models of gas emission behind the front of a
shock propagating with the velocity of
100\mbox{--}200~km~s$^{-1}$ gives the best agreement at the
abundance $12+\log(\textrm{O/H}) \simeq 8.15$ with the models of
purely collisional excitation or $12+\log(\textrm{O/H}) \simeq
8.35$ for the models that include the gas pre-ionization by a
shock.

\section*{ACKNOWLEDGMENTS}

This work was supported by the Russian Foundation for Basic Research (project
no.~07-02-00227). The work is based on the observational data obtained with the
6-m SAO telescope funded by the Ministry of Science of Russia (registration no.
01-43). When working on the paper, we used the NASA/IPAC Extragalactic Database
(NED) operated by the Jet Propulsion Laboratory of the California Institute of
Technology under contract with the National Aeronautics and Space
Administration (USA). We are grateful to the anonymous referee for helpful
remarks.

\bigskip
\centerline{REFERENCES}
\bigskip

\noindent V.~L.~Afanasiev and A.~V.~Moiseev, Pis'ma Astron. Zh. \textbf{31}, 269 (2005) [Astron. Lett. \textbf{31}, 194 (2005)].

\noindent  M.~G.~Allen, B.~A.~Groves, M.~A.~Dopita, et~al., Astrophys. J.~Suppl. Ser. \textbf{178}, 20 (2008).

\noindent  L.~H.~Aller, \emph{Physics of Thermal Gaseous Nebulae} (Reidel, Dordrecht, 1984).

\noindent  N.~V.~Asari, R.~Cid Fernandes, G.~Stasinska, et~al., Mon. Not. R.~Astron. Soc. \textbf{381}, 263 (2007).

\noindent  J.~Borissova, L.~Georgiev, M.~Rosado, et~al., Astron. Astrophys. \textbf{363}, 130 (2000).

\noindent  A.~Bullejos and M.~Rozado, Rev. Mex. Astron. Astrophys. \textbf{12}, 254 (2002).

\noindent  S.~Charlot and M.~Longhetti, Mon. Not. R.~Astron. Soc. \textbf{323}, 887 (2001).

\noindent  K.~T.~Chyzy, J.~Knapik, D.~J.~Bomans, et~al., Astron. Astrophys. \textbf{405}, 513 (2003).

\noindent  R.~Cid~Fernandes, N.~V.~Asari, L.~Sodre, et~al., Mon. Not. R.~Astron. Soc. \textbf{375}, 16 (2007).

\noindent  P.~A.~Crowther, L.~Drissen, J.~B.~Abbott, et~al., Astron. Astrophys. \textbf{404}, 483 (2003).

\noindent  S.~Demers, P.~Battinelli, and B.~Letarte, Astron. Astrophys. \textbf{424}, 125 (2004).

\noindent  M.~A.~Dopita and R.~S.~Sutherland, Astrophys. J. \textbf{455}, 468 (1995).

\noindent  M.~A.~Dopita and R.~S.~Sutherland, Astrophys. J.~Suppl. Ser. \textbf{102}, 161 (1996).

\noindent  M.~A.~Dopita, J.~Fischera, R.~S.~Sutherland, et~al., Astrophys. J.~Suppl. Ser. \textbf{167}, 177 (2006).

\noindent  V.~C.~Galarza, R.~A.~M.~Walterbos, and R.~Braun, Astron. J. \textbf{118}, 2775 (1999).

\noindent  D.~R.~Garnett, Astrophys. J. \textbf{363}, 142 (1990).

\noindent  A.~Gil~de~Paz, B.~F.~Madore, and O.~Pevunova, Astrophys. J.~Suppl. Ser. \textbf{147}, 29 (2003).

\noindent  A.~M.~Hidalgo-Gamez, Astron. Astrophys. \textbf{442}, 443 (2005).

\noindent  P.~Hodge and M.~G.~Lee, Publ. Astron. Soc. Pacific \textbf{102}, 26 (1990).

\noindent  D.~A.~Hunter, Astrophys. J.~\textbf{559}, 225 (2001).

\noindent  Y.~I.~Izotov, G.~Stasinska, G.~Meynet, et~al., Astron. Astrophys. \textbf{448}, 955 (2006).

\noindent  H.~Lee, M.~L.~McCall, R.~L.~Kingsburgh, et~al., Astron. J.~\textbf{125}, 146 (2003).

\noindent  J.~Lequeux, M.~Peimbert, J.~F.~Rayo, et~al., Astron. Astrophys. \textbf{80}, 155 (1979).

\noindent  A.~Leroy, A.~Bolatto, F.~Walter, and L.~Blitz, Astrophys. J.~\textbf{643}, 825 (2006).

\noindent  T.~A.~Lozinskaya and A.~V.~Moiseev, Mon. Not. R.~Astron. Soc. \textbf{381}, 26L (2007).

\noindent  T.~A.~Lozinskaya, A.~V.~Moiseev, N.~Yu.~Podorvanyuk, and A.~N.~Burenkov, Pis'ma Astron. Zh. \textbf{34}, 243 (2008) [Astron. Lett. \textbf{34}, 217 (2008)].

\noindent  P.~Massey and S.~Holmes, Astrophys. J.~Lett. \textbf{580}, L35 (2002).

\noindent  P.~Massey, T.~E.~Armandroff, and P.~S.~Conti, Astron. J. \textbf{103}, 1159 (1992).

\noindent  P.~Massey, K.~Olsen, P.~Hodge, G.~Jacoby, R.~McNeill, R.~Smith, and Sh.~Strong, Astron. J.~\textbf{133}, 2393 (2007).

\noindent  B.~E.~J.~Pagel, M.~G.~Edmunds, D.~E.~Blackwell, et~al., Mon. Not. R.~Astron. Soc. \textbf{189}, 95 (1979).

\noindent  L.~S.~Pilyugin and T.~X.~Thuan, Astrophys. J. \textbf{631}, 231 (2005).

\noindent  C.~J.~Ramsey, W.~R.~Milliams, R.~A.~Gruendl, et~al., Astrophys. J. \textbf{641}, 241 (2006).

\noindent  M.~G.~Richer, A.~Bullejos, J.~Borissova, et al., Astron. Astrophys. \textbf{370}, 34 (2001).

\noindent  M.~Rosado, M.~Valdez-Gutierrez, A.~Bullejos, et~al., ASP Conf. Ser. \textbf{282}, 50 (2002).

\noindent  P.~Royer, S.~J.~Smartt, J.~Manfroid, and J.~Vreux, Astron. Astrophys. \textbf{366}, L1 (2001).

\noindent  S.~Sakai, B.~F.~Madore, and W.~L.~Freedman, Astrophys. J. \textbf{511}, 671 (1999).

\noindent  J.~C.~Thurow and E.~M.~Wilcots, Astron. J. \textbf{129}, 745 (2005).

\noindent  W.~D.~Vacca, C.~D.~Sheehy, and J.~R.~Graham, Astrophys. J. \textbf{662}, 272 (2007).

\noindent  E.~M.~Wilcots and B.~W.~Miller, Astron. J. \textbf{116}, 2363 (1998).

\noindent  H.~Yang and E.~D.~Skillman, Astron. J. \textbf{106}, 1448 (1993).

\noindent  D.~B.~Zucker, Bull. Am. Astron. Soc. \textbf{32}, 1456 (2000).

\noindent  D.~B.~Zucker, Bull. Am. Astron. Soc. \textbf{34}, 1147 (2002).



\end{document}